\documentclass[acmsmall,screen]{acmart}
\AtBeginDocument{%
  }

\setcopyright{acmcopyright}
\copyrightyear{2025}
\acmYear{2025}
\acmDOI{XXXXXXX.XXXXXXX}

\usepackage{tabularx}
\begin{document}

\title{A Study of the Framework and Real-World Applications of Language Embedding for 3D Scene Understanding}


\author{Mahmoud Chick Zaouali}
\email{mahmoudchickzaouali@uvic.ca}
\orcid{0000-0002-5381-9736}

\author{Todd Charter}
\email{toddch@uvic.ca}
\orcid{0000-0001-5982-255X}

\author{Yehor Karpichev}
\email{ykarpichev@uvic.ca}
\orcid{0009-0008-5830-0723}

\author{Brandon Haworth}
\email{bhaworth@uvic.ca}
\orcid{0000-0001-8134-0047}

\author{Homayoun Najjaran}
\email{najjaran@uvic.ca}
\orcid{0000-0002-3550-225X}
\authornote{Corresponding Author}

\affiliation{%
  \institution{Faculty of Engineering and Computer Science, University of Victoria}
  \city{Victoria}
  \state{BC}
  \country{Canada}
}

\renewcommand{\shortauthors}{Zaouali et al.}

\begin{abstract}
Gaussian Splatting has rapidly emerged as a transformative technique for real-time 3D scene representation, offering a highly efficient and expressive alternative to Neural Radiance Fields (NeRF). Its ability to render complex scenes with high fidelity has enabled progress across domains such as scene reconstruction, robotics, and interactive content creation. More recently, the integration of Large Language Models (LLMs) and language embeddings into Gaussian Splatting pipelines has opened new possibilities for text-conditioned generation, editing, and semantic scene understanding. Despite these advances, a comprehensive overview of this emerging intersection has been lacking. This survey presents a structured review of current research efforts that combine language guidance with 3D Gaussian Splatting, detailing theoretical foundations, integration strategies, and real-world use cases. We highlight key limitations such as computational bottlenecks, generalizability, and the scarcity of semantically annotated 3D Gaussian data and outline open challenges and future directions for advancing language-guided 3D scene understanding using Gaussian Splatting.
\end{abstract}



\received{20 February 2007}
\received[revised]{12 March 2009}
\received[accepted]{5 June 2009}

\maketitle

\section{Introduction}

Driven by applications in robotics, autonomous navigation, and entertainment, as well as the growing need for immersive experiences in virtual and augmented reality, 3D scene reconstruction has emerged as a crucial area of research in computer vision and graphics. Novel view synthesis (NVS) techniques have advanced significantly in recent years, with two notable approaches gaining considerable attention in the research community: 3D Gaussian Splatting (3DGS) \cite{kerbl20233d} and Neural Radiance Fields (NeRF) \cite{mildenhall2021nerf}. 

Prior to NeRF, several neural implicit representations had already demonstrated the feasibility of encoding 3D geometry in continuous function spaces. Occupancy Networks \cite{mescheder2019occupancy} and DeepSDF \cite{park2019deepsdf}, introduced in 2018 and 2019 respectively, represented 3D shapes using neural networks to model either occupancy probabilities or signed distance fields. These approaches laid essential groundwork for subsequent volumetric scene representations.

Building on these ideas, NeRF introduced a paradigm shift by incorporating view-dependent radiance into the representation. Using three spatial coordinates ($x$, $y$, and $z$) to represent a point in 3D space and two angular coordinates ($\theta$ and $\phi$) to define the viewing direction relative to the scene, NeRF revolutionized implicit 3D scene encoding using deep fully-connected neural networks. This structure enables the model to generate both volume density and view-dependent RGB colors. Even with its remarkable rendering quality, NeRF is not suitable for real-time applications because of its long training and inference times \cite{muller2022instant}. To address NeRF’s performance limitations, Instant-NGP \cite{muller2022instant} introduced architectural improvements that enabled real-time training and rendering. However, despite these optimizations, its reliance on implicit scene representations continued to limit reconstruction flexibility and control.

In contrast, \citet{kerbl20233d} introduced 3D Gaussian Splatting as a flexible and explicit scene representation. Like many early NeRF-based methods, it begins with camera poses estimated via Structure-from-Motion (SfM) \cite{snavely2006photo}, using the resulting sparse point cloud to initialize a set of 3D Gaussians. Unlike point-based methods that rely on dense Multi-View Stereo reconstructions, 3DGS achieves high-quality results using only this sparse initialization. It then optimizes a differentiable volumetric representation where each Gaussian is projected to 2D using standard $\alpha$-blending \cite{mildenhall2021nerf}, enabling efficient and photorealistic rendering.

Parallel to these advances in 3D representation, the rise of Large Language Models (LLMs) and Vision-Language Models (VLMs) has reshaped the landscape of computer vision and Artificial Intelligence. LLMs have demonstrated remarkable capabilities in language understanding, generation, translation, and reasoning, serving as copilots in tasks ranging from writing to coding. At the same time, VLMs have redefined computer vision by enabling open-vocabulary recognition, zero-shot classification, segmentation, and grounding tasks that were traditionally constrained by closed, category-specific training. These vision-based models leverage aligned image-text representations to understand and label visual content in a more generalizable and semantically rich manner. Building on this foundation, Multimodal LLMs (MLLMs) have emerged that can process and reason across multiple data modalities such as language, vision, and audio, thereby mimicking human-like perception and interaction. 

However, while VLMs and MLLMs have significantly advanced 2D perception tasks, extending these capabilities to 3D remains a major challenge. Our world is inherently three-dimensional, and achieving spatial reasoning, object interaction, and scene-level understanding in 3D requires geometric consistency, multi-view alignment, and accurate camera estimation; problems that 2D models are not equipped to handle directly. Naively projecting 2D knowledge into 3D space often leads to ambiguity and a loss of both structural and semantic fidelity. For future embodied AI systems that interact with real-world environments, a deep and structured understanding of 3D scenes from scene-level layout to fine-grained object semantics is essential. Relying solely on 2D vision limits this potential. Meanwhile, VLMs and MLLMs have brought powerful world knowledge, compositional reasoning, and generalization capabilities that, if properly harnessed, can significantly improve 3D scene understanding.

With the rapid adoption of 3D Gaussian Splatting as a new standard for photorealistic and real-time 3D scene representation, and the growing trend of integrating foundation models into 3D pipelines, we find it timely and necessary to review this emerging intersection. In particular, a growing number of works have begun exploring the integration of VLMs and LLMs into 3D scene representations to tackle tasks like semantic scene understanding, grounding, captioning, and interaction \cite{ma2024llmsstep3dworld, ding2025world, HowtoEnableLLMwith3DCapacity}. However, none have comprehensively addressed the integration of language embeddings with 3DGS. Our goal is to provide the first in-depth perspective of this rapidly evolving direction.

This paper adopts a tutorial-style approach, aiming to guide the reader through the core components necessary to understand and evaluate the integration of language embeddings into 3D scene understanding. We begin by laying the groundwork with 3D Gaussian Splatting, a state-of-the-art method for real-time 3D scene representation, outlining its core pipeline, and distinguishing it from NeRF-based techniques. Next, we explore the evolution of language embedding methods from early word embeddings to modern LLMs and VLMs. Building on these foundations, we examine how language models are now being integrated with 3DGS to tackle complex scene understanding tasks. Throughout the paper, we highlight real-world applications, discuss current limitations, and identify open research directions to inform and inspire future advancements in this emerging field.

\section{Fundamentals of Gaussian Splatting}

3D Gaussian Splatting has emerged as an efficient and flexible approach for real-time 3D scene representation and rendering. Unlike volume-based neural representations such as NeRFs, which rely on computationally expensive volumetric ray marching, 3DGS represents a scene using a collection of parameterized 3D Gaussians. Each Gaussian is defined by its spatial position, opacity, shape (anisotropic covariance), and color information, making it an explicit and differentiable representation that can be rasterized efficiently \cite{kerbl20233d}.

\begin{figure*}[!htbp]
    \centering    \includegraphics[width=0.9\textwidth]{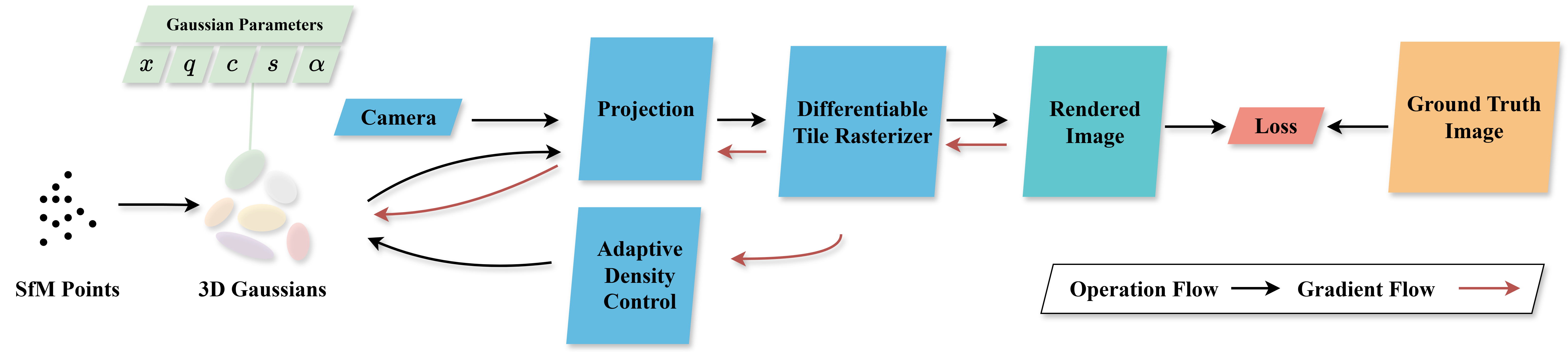}
\caption{Overview of the Gaussian Splatting Pipeline. The pipeline illustrates the key modules of 3D Gaussian Splatting. Gaussians are initialized from sparse SfM point clouds, then projected and rasterized to produce a rendered image, which is compared to ground truth using a loss function. Gradients flow backward to update Gaussian parameters. Adaptive Density Control adds or removes Gaussians based on scene geometry \cite{kerbl20233d}.}
\Description{Gaussian Splatting Pipeline}
\label{fig:Gaussian_Splatting_Pipeline}
\end{figure*}

The approach builds upon traditional Structure-from-Motion techniques, initializing a sparse set of 3D Gaussians from point clouds produced by SfM-based camera calibration. Unlike many point-based methods that require dense Multi-View Stereo reconstructions, Gaussian Splatting can generate high-quality novel view synthesis using only sparse point clouds as input. This process is illustrated in Fig. \ref{fig:Gaussian_Splatting_Pipeline}. 

\subsection{Gaussians as a Scene Geometry }
Each point in the scene is treated as a full 3D Gaussian characterized by a mean position $\mu$, a covariance matrix $\Sigma$, an opacity value $\alpha$, and an appearance model parameterized using Spherical Harmonics (SH) where the primitive is defined as: 

\begin{equation}
G(x) = e^{-\frac{1}{2} (x - \mu)^T \Sigma^{-1} (x - \mu)}
\end{equation}

For rendering, each 3D Gaussian must be projected onto 2D screen space, where it is represented as an elliptical splat. This transformation follows the approach introduced in \cite{zwicker2001ewa}, where the projected covariance $\Sigma'$ in camera coordinates is computed as:

\begin{equation} 
\Sigma' = J W \Sigma W^T J^T 
\end{equation}

Here, $W$ is the viewing transformation matrix, and $J$ represents the Jacobian of the affine approximation of the projective transformation. To obtain the final 2D covariance matrix, the third row and column of $\Sigma'$ are removed, resulting in a 2×2 variance matrix. This step produces a representation equivalent to previous 2D point-based methods that used planar discs with surface normals. However, unlike these prior approaches, 3DGS does not require explicit surface normal estimation, as the full 3D covariance matrix $\Sigma$ inherently encodes the anisotropic shape of the splats.

\subsection{Formulating the Problem for Efficient Optimization}

A key challenge in optimizing the 3D Gaussian representation is ensuring that the covariance matrix $\Sigma$ remains positive semi-definite, as directly optimizing matrix components can lead to numerical instability. Instead, 3DGS parameterizes $\Sigma$ as a composition of a scaling matrix $S$ and a rotation matrix $R$ to avoid invalid covariance matrices due to gradient-based updates.

\begin{equation} 
\Sigma = R S S^T R^T 
\end{equation}

where $S$ is a diagonal scaling matrix that defines the extent of the Gaussian along its principal axes and $R$ is a rotation matrix that determines its orientation in world space. 

To ensure stable optimization, $S$ and $R$ are stored separately, where the scaling values are represented as a 3D vector $s$ and the rotation is parameterized as a quaternion $q$, which is normalized to ensure a valid unit quaternion when converting it into a rotation matrix. Gradient-based optimization is performed on all Gaussian parameters, including position, opacity, and spherical harmonic coefficients, using explicit gradient computations instead of relying on automatic differentiation. This reduces computational overhead while maintaining differentiability across all steps. 

Optimization of 3DGS is performed through gradient-based updates, iteratively refining the scene representation by rendering and comparing synthesized images to training views. Due to 3D-to-2D projection ambiguities, the optimization process must not only refine existing Gaussians but also create, reposition, or remove them as needed. Covariance parameters play a crucial role in ensuring a compact representation, as large anisotropic Gaussians can efficiently model homogeneous regions. 

Stochastic Gradient Descent (SGD) techniques are used, leveraging GPU-accelerated frameworks with custom CUDA kernels to improve efficiency. Rasterization is the primary computational bottleneck, making an optimized differentiable rasterizer essential for fast training. To ensure stable optimization, opacity is constrained using a sigmoid function, while Gaussian scale parameters are updated with an exponential activation to maintain numerical stability. The loss function combines L1 loss with a D-SSIM term, balancing pixel-level accuracy with structural similarity for high-quality reconstruction.

To maintain an accurate scene representation, 3DGS dynamically adjusts the number and density of Gaussians throughout optimization. The process begins with an initial sparse set of points obtained from SfM, which are progressively refined by analyzing positional gradients. Areas with high positional gradients are strong candidates for densification, as they indicate regions where reconstruction is incomplete or where Gaussians cover overly large regions. In under-reconstructed areas where geometry is missing, new Gaussians are introduced by duplicating existing ones and shifting them along positional gradients to improve coverage. In contrast, in over-reconstructed regions where Gaussians cover large areas with high variance, existing Gaussians are split into smaller ones to better capture finer details.

To regulate the overall number of Gaussians, an opacity regularization strategy is applied. Periodically reducing opacity values allows the optimization process to reinforce necessary Gaussians while naturally eliminating redundant ones. Unlike volumetric approaches that rely on space compaction or warping techniques, 3DGS maintains a fully Euclidean representation, avoiding additional transformations required in other methods and simplifying the optimization process.

\subsection{The Differentiable Tile-Based Rasterizer}

A key component of 3DGS is its efficient, fully differentiable rasterization pipeline, enabling real-time rendering and gradient-based optimization. Instead of costly per-pixel sorting, 3DGS employs a tile-based rasterizer that streamlines sorting and blending, allowing scalable rendering of Gaussian splats.

The rasterization pipeline begins with frustum culling, discarding Gaussians outside the camera’s view by checking whether their  $99\%$ confidence intervals intersect the frustum. Guard band culling removes Gaussians near the camera plane to prevent numerical instability during 2D covariance computation. The remaining Gaussians are projected onto the screen space using view and projection matrices, transforming their 3D mean $\mu$ and covariance $\Sigma$ into 2D ($\mu'$ and $\Sigma'$).

A bounding rectangle determines screen overlap. If present, color is computed using Spherical Harmonics ($SH$), capturing view-dependent appearance. To enable correct blending, Gaussians are sorted by depth using a tile-based method, avoiding expensive per-pixel sorting. Efficient GPU-based sorting, such as Radix Sort \cite{10.1145/1854273.1854344}, processes millions of Gaussians quickly.

The image is divided into fixed-size tiles, with each CUDA thread block handling one tile. Shared memory optimizes Gaussian fetching and accumulation. $\alpha$-blending composites splats front-to-back, with early termination when full opacity ($\alpha = 1$) is reached to reduce computation. Opacity is computed with a differentiable function:
\begin{equation} \alpha_i = o_i \cdot \exp\left(-\frac{1}{2} (p' - \mu')^T \Sigma'^{-1} (p' - \mu')\right) \end{equation}

where  $\alpha_i$ is the opacity of the $i$-th Gaussian, $o_i$ is the opacity of the $i$-th is the Gaussian’s opacity parameter, $p'$ is the pixel position, $\mu'$ is the 2D mean, and $\Sigma'$ is the 2D covariance matrix. Gaussians with very low opacity ($\alpha<1/255$) are discarded. If transmittance $T_i$ drops below a threshold, rendering stops early. Final pixel color is computed using the differentiable volume rendering equation:

\begin{equation} C = \sum_{i=1}^{N} T_i \alpha_i c_i \end{equation}

where, $T_i = \prod_{j=1}^{i-1} (1 - \alpha_j)$ is the transmittance, $\alpha_i = 1 - \exp(-\sigma_i \delta_i)$ is the opacity derived from the Gaussian's density ($\sigma_i$) and depth ($\delta_i$), $c_i$ is the color of the $i$-th Gaussian.

A key strength of this approach is its ability to propagate gradients efficiently without restricting how many Gaussian splats contribute to each pixel. Unlike earlier rendering methods that imposed a fixed limit on the number of blended splats during training, 3DGS supports an arbitrary number of overlapping Gaussians. This flexibility enables the model to learn effectively across scenes with varying depth complexity and eliminates the need for manual tuning of hyperparameters related to blending limits.

During backpropagation, the sorted list from the forward pass is reused. Instead of storing all intermediate opacities, only the final accumulated opacity is saved. Intermediate blending weights are recovered via back-to-front traversal, enabling accurate gradient computation for each Gaussian.

\section{Language Embeddings and LLMs for 3D Scene Understanding}

\begin{figure*}[!htbp]
    \centering    \includegraphics[width=\textwidth]{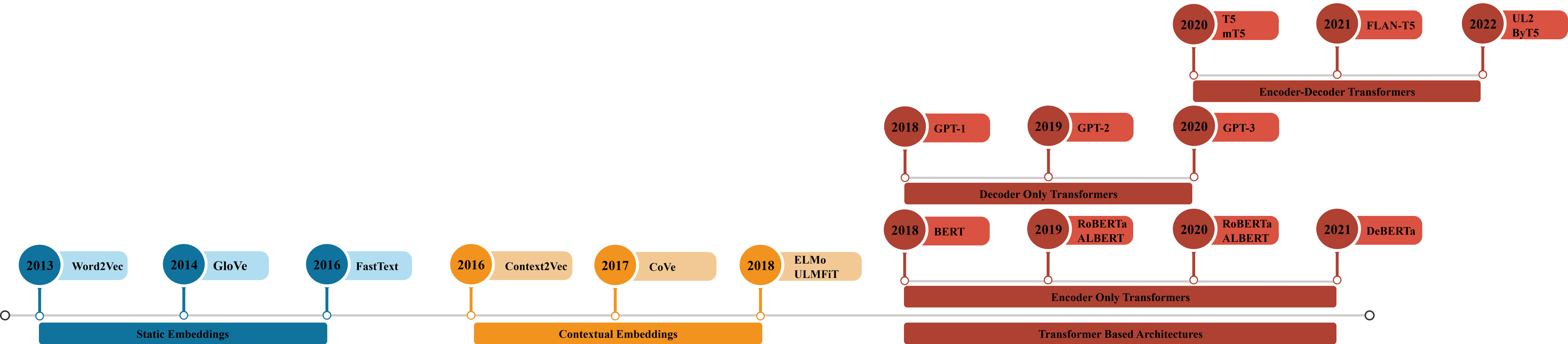}
\caption{Evolution of Language Embedding Techniques and Transformer-Based Architectures.
The timeline traces the shift from static to contextual embeddings, and the rise of transformer-based models categorized as decoder-only, encoder-only, and encoder-decoder architectures.}
\Description{Timeline of Language Embedding Techniques and the Emergence of Transformer-Based Language Models}
\label{fig:language_embedding_timeline}
\end{figure*}

Understanding complex 3D scenes through natural language requires powerful language representations that capture both semantic richness and contextual nuance. This section traces the evolution of language embeddings from classical models that offered static representations, to contextual models that leverage deep Transformer architectures for dynamic, context-aware understanding. We then explore multimodal embeddings, which align language with visual signals, beginning with 2D vision-language models and progressing toward large-scale vision foundation models that serve as robust priors for image and scene interpretation. Finally, we connect these advances to 3D scene understanding, examining how language-grounded models and emerging 3D MLLMs integrate linguistic and spatial reasoning to enable holistic, grounded interpretations of complex environments. This section sets the stage for understanding how modern AI systems fuse textual and geometric information to interact meaningfully with the 3D world. The progression of these models is shown in Fig. \ref{fig:language_embedding_timeline}.

\subsection{Word Embeddings}

Early word embedding techniques such as Word2Vec \cite{mikolov2013efficient}, GloVe \cite{pennington-etal-2014-glove}, and FastText \cite{bojanowski2017enriching} played a pivotal role in representing semantic meaning in vector space. Word2Vec learns dense word representations by leveraging local context through two training objectives: Continuous Bag-of-Words (CBOW), which predicts a target word from surrounding words, and Skip-Gram, which predicts context words from a target word. Despite their efficiency, these models primarily capture short-range dependencies. GloVe, on the other hand, introduced a global perspective by using word co-occurrence statistics across the entire corpus. Instead of predicting words based on immediate context, GloVe learns word vectors that reflect the relative frequency of word pairs, capturing broader semantic relationships. However, both Word2Vec and GloVe treat words as atomic units, limiting their ability to represent rare or morphologically complex words. FastText addresses this by modeling words as a combination of character-level n-grams, enabling the model to compose embeddings for unseen or rare words based on subword patterns. This subword modeling makes FastText especially effective for morphologically rich languages. While these approaches significantly improved semantic representations, they produce static embeddings that do not adapt to different linguistic contexts, a limitation that would later be addressed by contextual models based on transformers.

Early attempts to incorporate context into word representations include Context2Vec \cite{melamud-etal-2016-context2vec}, which used a bidirectional LSTM to model sentence-level semantics, and CoVe \cite{McCann2017LearnedIT}, which extended this idea using neural machine translation. These models generated dynamic embeddings based on surrounding text, addressing word sense disambiguation; for example, distinguishing between "bat" as a flying mammal or a piece of sports equipment. Often, these contextual embeddings were combined with static vectors like GloVe to balance global semantics with contextual nuance.

\subsection{Large Language Models}

The introduction of pretrained language models (PLMs) marked a major shift. ELMo \cite{peters-etal-2018-deep}, based on a deep bidirectional language model, generated embeddings by jointly modeling both left and right contexts. Its multi-layer architecture encoded syntactic and semantic features at different depths, outperforming earlier models on a variety of NLP benchmarks. ULMFiT \cite{howard2018universal} further advanced transferability by enabling task-specific fine-tuning through techniques like discriminative learning rates and gradual unfreezing, reducing reliance on large labeled datasets.

Overall, these models laid the foundation for modern transformer-based architectures, offering dynamic, task-adaptable word representations that significantly improve performance in downstream language tasks.

The introduction of the Transformer architecture \cite{vaswani2017attention} revolutionized natural language processing by enabling parallelized learning and more effective modeling of long-range dependencies. Unlike earlier recurrent models, Transformers rely entirely on self-attention mechanisms, allowing them to capture contextual relationships across an entire sequence efficiently.

This architectural shift laid the foundation for modern LLMs such as BERT \cite{devlin-etal-2019-bert}, GPT \cite{radford2018improving}, and T5 \cite{raffel2020exploring}, which now form the backbone of most state-of-the-art NLP systems. These models are pre-trained on large-scale corpora and fine-tuned for a variety of downstream tasks, offering contextualized word representations that adapt to how words are used in different sentences.

Transformer-based models differ in how they are structured and applied: encoder-only models like BERT are optimized for language understanding, decoder-only models like GPT for text generation, and encoder-decoder models like T5 for sequence-to-sequence tasks such as translation and summarization. These variations reflect different pretraining objectives and have led to a diverse set of large-scale models adapted for downstream tasks. In the following sections, we explore representative examples from each category, starting with GPT and BERT.

\textbf{The Generative Pretrained Transformer (GPT)} \cite{radford2018improving} represents a pivotal development in language modeling, showcasing the scalability and flexibility of the decoder-only Transformer architecture \cite{vaswani2017attention}. Introduced by OpenAI in 2018, GPT-1 leveraged a two-phase training paradigm: unsupervised pretraining on large-scale unlabeled text corpora, followed by supervised fine-tuning on specific downstream tasks. This approach enabled the model to learn rich, universal language representations without requiring extensive labeled data.

Building on this foundation, GPT-2 \cite{Radford2019LanguageMA} scaled the architecture significantly (to 1.5B parameters) and introduced a new training perspective, unsupervised multitask learning. Unlike its predecessor, GPT-2 required no explicit task-specific fine-tuning. Instead, it learned to perform a variety of NLP tasks such as translation, summarization, and question answering by modeling them as conditional language generation problems. That is, given an input prompt and optional task description, GPT-2 generates the output directly, unifying multiple tasks under a single objective. This formulation enabled zero-shot and few-shot generalization, where the model could perform new tasks with minimal or no task-specific training data.

Despite these advances, GPT-2 still faced limitations in capacity and generalization. It underfit its training data (WebText) and lagged behind fine-tuned models in some benchmarks. These constraints motivated the exploration of scaling laws, most notably the work of \citet{kaplan2020scaling} and \citet{hoffmann2022training}, which provided empirical evidence that performance improves predictably with increased model size, dataset size, and compute.

This insight led to the release of GPT-3 \cite{NEURIPS2020_1457c0d6}, a model with 175B parameters, over 10× larger than GPT-2, while maintaining the same decoder-only architecture. GPT-3 introduced the concept of in-context learning (ICL), where the model learns to perform tasks by conditioning on examples provided in the input prompt, without updating its parameters. OpenAI described this as a form of meta-learning, where the model learns a broad range of skills during pretraining and applies them adaptively at inference time.

GPT-3 demonstrated strong performance across standard NLP tasks, reasoning challenges, and domain-specific applications, often rivaling or surpassing fine-tuned models. Its success marked a key transition from conventional pre-trained language models to general-purpose LLMs with emergent capabilities such as instruction following, common-sense reasoning, and flexible adaptation to novel tasks; all driven by scale, data, and architecture \cite{zhao2024surveylargelanguagemodels, 10.1145/3729218}.

\textbf{Bidirectional Encoder Representations from Transformers (BERT)} \cite{devlin-etal-2019-bert} is a foundational encoder-only model designed for natural language understanding (NLU) tasks \cite{lin2022survey}. Unlike unidirectional models such as GPT, BERT leverages the Transformer encoder to learn deep, bidirectional representations by jointly conditioning on both left and right context. This enables a more comprehensive understanding of words in context. 

Pre-trained using large-scale unlabeled text, BERT adopts two self-supervised objectives: Masked Language Modeling (MLM), where tokens are randomly masked and predicted based on surrounding words, and Next Sentence Prediction (NSP), which helps model inter-sentence relationships. These objectives enable BERT to capture both token-level and sentence-level semantics, facilitating strong performance on a range of downstream tasks via simple fine-tuning.

BERT’s success sparked a wave of optimized variants. RoBERTa \cite{liu2019roberta} improved training efficiency and performance by removing NSP, using dynamic masking, and scaling up data and training time. ALBERT \cite{lan2020albertlitebertselfsupervised} reduced model size through parameter sharing and factorized embeddings, while introducing Sentence Order Prediction (SOP) to better model discourse-level coherence. DistilBERT \cite{sanh2020distilbertdistilledversionbert} applied knowledge distillation to compress BERT into a smaller, faster model with minimal performance loss, making it suitable for real-time applications.

Further extensions such as ELECTRA \cite{clark2020electrapretrainingtextencoders}, which replaces MLM with a more efficient replaced-token detection objective, and DeBERTa \cite{he2021debertadecodingenhancedbertdisentangled}, which introduces disentangled attention and improved positional encoding, illustrate the continued innovation in BERT-style architectures. Together, these models demonstrate the adaptability of encoder-only Transformers for scalable, high-performance NLP systems.

\section{Vision-Language, Diffusion and Foundation Models}
Recent progress in 3D scene understanding has been fueled by advances in three major categories: vision-language models that align text with visual inputs, diffusion models that enable text-driven image and scene generation, and vision foundation models trained on massive datasets to generalize across visual tasks. In this section, we examine each of these model types and their growing influence on 3D perception, interaction, and synthesis.

\subsection{2D Vision-Language Models}
Vision-Language Models (VLMs) form a foundational class of multimodal models that learn joint representations across image (or video) and text modalities. Originally designed for 2D tasks, these models are now being adapted to support language-driven perception and interaction in 3D environments. Most VLMs adopt transformer-based architectures with distinct visual and textual encoders, whose outputs are aligned using contrastive learning or cross-attention mechanisms. Depending on the training objectives, VLMs can be geared toward discriminative tasks (e.g., classification, retrieval) or generative tasks (e.g., captioning, Visual Question Answering, synthesis) \cite{ma2024llmsstep3dworld}.

\textbf{Contrastive Vision-Language Models:} Contrastive VLMs aim to learn semantically aligned embeddings across modalities by encouraging matched image–text pairs to be close in the shared latent space while separating mismatched pairs. Prominent examples such as CLIP \cite{pmlr-v139-radford21a} and ALIGN \cite{pmlr-v139-jia21b} were trained on massive web-scale image-caption datasets, enabling them to generalize remarkably well to unseen tasks through zero-shot inference. This training paradigm enables tasks like open-vocabulary classification and cross-modal retrieval by comparing query images to candidate textual descriptions in the embedding space without the need for additional fine-tuning.

Although initially introduced for image classification, these models have demonstrated versatility across tasks such as object detection, image segmentation, document analysis, and even video recognition. A common strategy for adapting contrastive models to new domains involves vision-language knowledge distillation, where the general knowledge of a pretrained model like CLIP (acting as a teacher) is transferred to smaller or task-specific student models \cite{ma2024llmsstep3dworld}.

While CLIP and ALIGN achieve strong semantic grounding, they rely heavily on global representations and lack fine-grained localization, which poses challenges for tasks requiring spatial understanding, an important consideration when extending such models to 3D scene understanding.

\textbf{Generative Vision-Language Models:} Other VLMs extend the scope of multimodal learning by enabling image-conditioned text generation, visual reasoning, and multimodal interaction. Models such as SimVLM \cite{wang2022simvlmsimplevisuallanguage}, BLIP \cite{pmlr-v162-li22n} and OFA \cite{pmlr-v162-wang22al} focus on image-to-text generation, excelling in tasks like image captioning and Visual Question Answering (VQA). More advanced models, including BLIP-2 \cite{pmlr-v202-li23q}, Flamingo \cite{10.5555/3600270.3601993}, and LLaVA \cite{liu2023visual}, incorporate multi-turn dialogue and contextual reasoning, enabling them to generate responses that are conditioned on both the image content and prior interactions \cite{ma2024llmsstep3dworld}.

While these transformer-based generative models enable powerful multimodal reasoning and dialogue, the rise of diffusion models has pushed the frontier further in high-fidelity image and scene generation, as we detail in the next section.

\subsection{Diffusion Models}
In parallel with VLMs, diffusion-based generative frameworks have emerged as a powerful tool for multimodal synthesis, particularly in generating visual content conditioned on textual prompts. Below, we explore their applications across image, video, and 3D domains.

\textbf{Diffusion Models for Text-to-Image Synthesis:}
The rise of diffusion models has driven significant research interest in tackling the challenge of generating images directly from textual descriptions. Diffusion models are a family of deep probabilistic generative models that have become state-of-the art for generaing high quality image samples, surpassing Generative Adversarial Networks (GANs) \cite{10.5555/2969033.2969125} in many applications. These models operate by progressively corrupting the data through the injection of noise in a forward diffusion process and then learning to reverse this process to generate realistic samples \cite{10.5555/3495724.3496298}. Traditional diffusion models operate directly in pixel space, requiring significant computational resources for optimization and resulting in slow inference due to their sequential evaluation. This inefficiency arises because image synthesis is decomposed into multiple iterative denoising steps, making high-resolution image generation computationally expensive. To address these limitations, \citet{rombach2022high} proposed Latent Diffusion Models (LDMs), which apply the denoising process in the latent space of pretrained autoencoders, significantly reducing computational demands while preserving high visual fidelity. Furthermore, by incorporating cross-attention layers, LDMs enable text conditioning, allowing textual descriptions to guide the image generation process. A related approach, VQ-Diffusion, introduced by \citet{Gu_2022_CVPR}, follows a similar latent-space modeling strategy but leverages a vector quantized variational autoencoder (VQ-VAE) to structure the latent representations differently for text-to-image synthesis. Beyond text-to-image synthesis, this approach has also been adapted for a range of other generative tasks, including video generation.

\textbf{Diffusion Models for Video Generation:}
\citet{NEURIPS2022_39235c56} introduced the first text-conditioned video generation model based on diffusion, demonstrating promising results. Their work was later extended by Imagen Video \cite{ho2022imagenvideohighdefinition}, which built upon this foundation to further advance video synthesis. By leveraging a simple yet effective architecture consisting of a frozen T5 text encoder, a base video diffusion model, and interleaved spatial and temporal super-resolution diffusion models, they demonstrated that the transition from text-to-image generation to video synthesis is feasible. 

\textbf{Diffusion Models for for Text-to-3D and Scene Editing:}
DreamFusion \cite{poole2022dreamfusiontextto3dusing2d} builds upon recent advancement in text-to-image synthesis by leveraging a pretrained 2D text-to-image diffusion model as a prior where they optimizes a NeRF representation through probability density distillation, enabling the synthesis of 3D scenes without requiring 3D training data. Similarly, MAV3D \cite{10.5555/3618408.3619731} extends this approach to dynamic 3D scene generation, making it the first method to synthesize 4D neural representations from text. Like DreamFusion, MAV3D employs NeRF as a scene representation, but optimizes it not only for scene appearance and density but also for motion consistency, leveraging pretrained diffusion-based models trained on text-image pairs and unlabeled videos. This enables the synthesis of dynamic scenes that can be viewed from any camera angle and composited into 3D environments. \citet{hertz2022prompttopromptimageeditingcross}, \citet{brooks2023instructpix2pix}, and \citet{mokady2023null} apply text-to-image diffusion models to modify existing images based on natural language instructions, while \citet{haque2023instruct} and \citet{10.1145/3610548.3618190} extend this approach to text-driven 3D scene editing.

\subsection{Vision Foundation Models}

Vision Foundation Models (VFMs) were inspired by the success of LLMs, aiming to build large-scale architectures trained on massive datasets to generalize across vision tasks. Early models relied solely on visual inputs, such as the Vision Transformer (ViT), which introduced self-attention for classification and retrieval. The Swin Transformer \cite{liu2021swin} improved on this by enabling efficient processing of high-resolution images, making it effective for object detection and segmentation. Self-supervised learning approaches like MAE, BEIT, and CAE leveraged masked modeling, where missing image regions were reconstructed to enhance spatial and semantic understanding. VideoMAE \cite{tong2022videomae} extended this masked modeling strategy to video, introducing high-ratio masking for more effective action classification and detection \cite{sun2024surveyreasoningfoundationmodels}.

DINO \cite{caron2021emerging} is a self-supervised learning framework that trains a vision transformer through self-distillation, where a student network is trained to mimic the output of a teacher network using different augmented views of the same image, without relying on any labels or human supervision. This training strategy encourages the model to focus on semantically meaningful regions, often corresponding to prominent objects, allowing it to distinguish foreground objects from background context without using any pixel-level supervision. 

DINOv2 \cite{oquab2023dinov2} builds on the DINO framework by scaling self-supervised learning with a larger ViT architecture and a curated dataset of 142 million images, resulting in high-quality, general-purpose visual features. The model learns both image-level and pixel-level representations that transfer effectively across a wide range of tasks, including classification, segmentation, and depth estimation without the need for fine-tuning. DINOv2 outperforms many existing self-supervised and weakly supervised models, such as CLIP and OpenCLIP \cite{cherti2023reproducible}, establishing itself as a strong backbone for versatile computer vision applications.

The Segment Anything Model (SAM) \cite{kirillov2023segment} introduced prompt-based segmentation with zero-shot generalization across various image segmentation challenges,  and its successor, SAM 2\cite{ravi2024sam}, extends this capability to videos through a unified architecture equipped with memory mechanisms for spatio-temporal segmentation, offering significantly improved accuracy and efficiency across both image and video domains. 

Another emerging trend involves combining individual foundation models to integrate their knowledge and tackle complex vision tasks, a strategy known in the literature as Model Fusion. As an example, SAM-Track \cite{cheng2023segment} combined Grounding DINO \cite{liu2024grounding} and DeAOT \cite{yang2022decoupling} with SAM to enable efficient object tracking and segmentation throughout a video sequence by incorporating interactive prompts such as click, box and text inputs, in the first frame to direct the segmentation task \cite{sun2024surveyreasoningfoundationmodels}. 

\section{Language Embeddings for 3D scene understanding}

\begin{figure*}[!htbp]
    \centering    \includegraphics[width=0.9\textwidth]{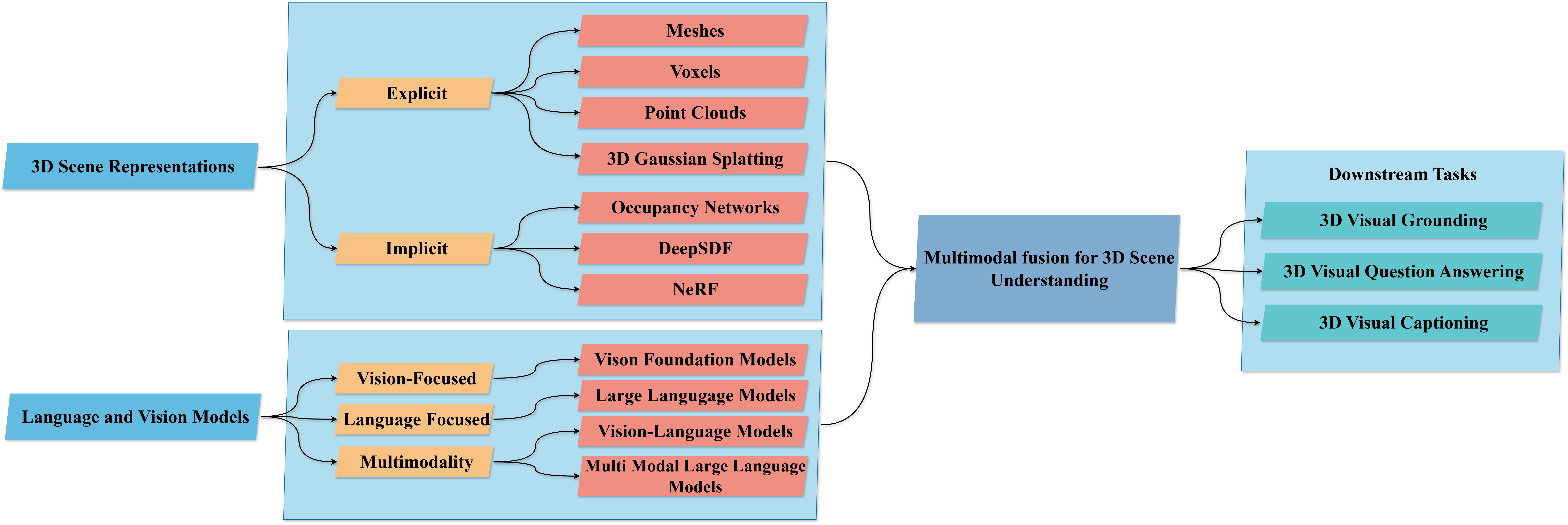}
\caption{Integration of 3D Scene Representations and Multimodal Language and Vision Models for 3D Scene Understanding. This diagram shows how explicit and implicit 3D representations are fused with vision, language, and multimodal models to enable downstream 3D scene understanding tasks.}
\Description{Integration of 3D Scene Representations and Multimodal Language-Vision Models for 3D Scene Understanding}
\label{fig:3d_scene_fusion}
\end{figure*}

3D scene understanding requires the integration of geometric representations with powerful language and vision models. Fig. \ref{fig:3d_scene_fusion} illustrates how multimodal fusion enables downstream tasks such as visual grounding, question answering, and captioning. The following subsections explore language-grounded understanding of 3D environments and recent advancements in 3D MLLMs.

\subsection{Language-Driven 3D Scene Understanding}

Language-driven 3D scene understanding focuses on integrating textual information with 3D environments, enabling models to comprehend and interact with scenes through natural language. One key task in this domain is 3D visual grounding, which involves identifying specific objects within a 3D scene based on textual queries. Models achieve this by detecting objects through 3D bounding boxes or segmentation masks, leveraging language cues to establish correspondence between descriptions and spatial entities \cite{ScanRefer, SceneVerse, ChenLanguageConditioned, zhu2023object2sceneputtingobjectscontext, ReferIt3D, huang2022multi, zhang2023multi3drefer, wang-etal-2023-3drp, Zhao_2021_ICCV, wang2023distilling, unal2024four}.

The second task is 3D dense captioning, which generates detailed and contextually relevant textual descriptions of 3D scenes. This process requires models to localize and describe multiple objects in a scene with high granularity, ensuring that the textual output captures the necessary semantic details \cite{chen2020scan2capcontextawaredensecaptioning, yuan2022x, JiaoMORE, chen2023end, 10.1109/TPAMI.2024.3387838}. 

The third task is 3D question answering, where models generate language-based responses to questions about a 3D environment. This involves reasoning about object attributes, spatial relationships, and general understanding of the scene to provide accurate answers \cite{azuma2022scanqa, parelli2023clip, ma2022sqa3d}.

Early research in language-driven scene understanding primarily focused on improving individual frameworks \cite{10.1007/978-3-031-20059-5_24, Zhao_2021_ICCV} or task-specific modules \cite{ChenLanguageConditioned, zhu2023object2sceneputtingobjectscontext}, which led to limited generalization performance across different tasks\cite{lyu2024mmscan}. To address this, some approaches explored task unification \cite{9879358, chen2022d}, integrating multiple tasks in the language of the scene to improve overall performance. Others introduced pre-training strategies \cite{zhu20233d, 10203209} to improve scene-language alignment and develop more adaptable models \cite{lyu2024mmscan}. Despite these advancements, many existing methods remain constrained by their task-specific architectures, limiting their flexibility for broader applications and downstream interactions.

\subsection{3D Multimodal LLMs}

Early developments in  3D MLLMs primarily focused on object-level understanding, utilizing abundant 3D-text data and relatively simpler architectures to establish fundamental scene comprehension \cite{10.1007/978-3-031-72698-9_8, guo2023pointbindpointllmaligning, 10.1007/978-3-031-72775-7_13, 10657856}. PointLLM \cite{10.1007/978-3-031-72698-9_8} follows this approach by directly mapping point-cloud data into an LLM’s embedding space, enabling language-based reasoning about individual objects without leveraging additional modalities. As research progressed, newer models sought to enhance scene-level comprehension by integrating multiple sensor inputs. Point-LLM \cite{guo2023pointbindpointllmaligning} and ImageBind-LLM \cite{han2023imagebindllmmultimodalityinstructiontuning} exemplify this shift by creating a joint embedding space that aligns 3D point clouds, images, audio, and text, enabling cross-modal understanding and richer spatial reasoning.

Although these multimodal models improved spatial reasoning, they remained largely object-centric. One of the first models to extend this to scene-level reasoning was 3D-LLM \cite{10.5555/3666122.3667022}, which leveraged pre-trained 2D VLM features while incorporating positional embeddings and location tokens to enhance spatial representation. However, its reliance on 2D encoders limited its ability to fully capture the complexities of 3D spatial structures and object relationships. 

Building on 3D-LLM's scene-level capabilities, Chat-3D \cite{wang2023chat3ddataefficientlytuninglarge} and LL3DA \cite{10656186}, sought to improve object-centric interactions by implementing preselection mechanisms, allowing for better alignment between 3D data and textual descriptions. Chat-3D also addressed the challenge of limited 3D-text data availability through a pre-alignment phase, ensuring more effective scene-text mapping. However, its architectural constraints limit interactions to predefined objects, reducing its flexibility in more generalized scene comprehension tasks.

Beyond explicit 3D-text alignment, a different line of research explores reasoning-driven scene understanding. The Embodied Generalist model \cite{huang2023embodied} emphasizes the core principles of situation understanding and situated reasoning \cite{ma2022sqa3d}, which are fundamental to embodied scene comprehension. By incorporating multimodal inputs, it enhances 3D scene understanding by enabling reasoning about spatial relationships and interactions within an environment. This approach bridges the gap between perception and action, ensuring a more comprehensive understanding of 3D spaces beyond passive recognition.

Despite these advancements, current 3D MLLMs struggle with precise object referencing and grounding, limiting their effectiveness in handling complex spatial reasoning tasks beyond simple object-level interactions. 

\section{Methods For Integrating Language Models With Gaussian Splatting}

\begin{figure*}[!htbp]
    \centering    \includegraphics[width=0.9\textwidth]{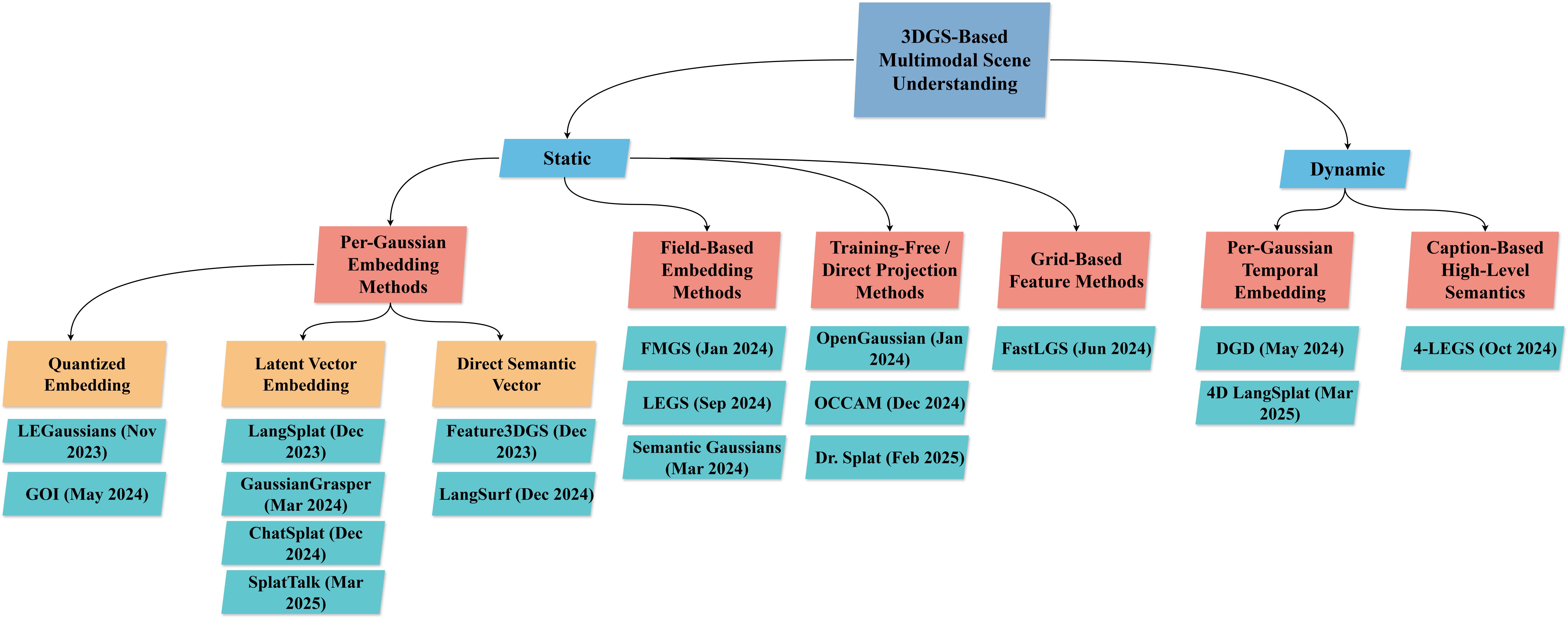}
\caption{Classification of 3DGS Methods for Vision-Language Scene Understanding. This figure categorizes 3DGS methods into static and dynamic groups based on their semantic embedding strategies, including per-Gaussian, field-based, training-free, grid-based, temporal, and caption-guided approaches. Representative methods and release dates are shown under each category.}
\Description{Classification of 3DGS Methods for Vision-Language Scene Understanding}
\label{fig:fig_3dgs_language_methods}
\end{figure*}

Recent work has extended Gaussian Splatting beyond geometry and appearance to incorporate language-driven semantics. As shown in Fig. \ref{fig:fig_3dgs_language_methods}, these methods can be broadly categorized based on whether they operate on static or dynamic scenes. The following subsections review approaches that embed language features into 3D Gaussians for both static environments and time-varying, dynamic settings.

\subsection{Static Scenes}

\citet{shi2024language} pioneered integrating language features into 3D Gaussian representations for open-vocabulary scene understanding. To avoid the memory and performance issues of directly embedding high-dimensional CLIP features, they proposed a language feature quantization method. This approach learns a discrete set of basis vectors (a codebook), representing each semantic feature by an index rather than storing full embeddings. This compresses the data and reduces noise by leveraging redundancy in scene semantics. To handle inconsistencies from multi-view observations, a learned uncertainty-guided smoothing mechanism is introduced, where each Gaussian stores a semantic uncertainty value that controls a positional MLP, promoting spatial consistency.

Building on similar goals, LangSplat  \cite{qin2024langsplat} takes a different approach. Rather than quantizing features, it encodes CLIP-extracted language features directly into 3D Gaussians. These features are obtained from image patches across views and compressed into a scene-specific latent space using a language autoencoder. Each Gaussian stores a compact latent vector, which is decoded during rendering. To enhance semantic precision, LangSplat incorporates hierarchical segmentation from SAM, assigning language features to exact object regions. This strategy reduces memory usage, speeds processing, and outperforms previous CLIP-supervised NeRF methods like LERF \cite{kerr2023lerf}.

Feature 3DGS \cite{zhou2024feature} leverages CLIP-LSeg and SAM to embed distilled semantic features directly into 3D Gaussians. Each Gaussian contains both its standard parameters and a semantic vector, enabling language-guided querying and editing of the scene. Language prompts are encoded using CLIP’s ViT-B/32, and cosine similarity is calculated between the prompt and each Gaussian’s semantic embedding. Softmax is applied to produce activation probabilities per category. This allows for semantic region selection via soft, hard, or hybrid strategies. Selected Gaussians can then be edited by changing opacity or color to enable fine-grained, prompt-driven tasks like segmentation, object deletion, and appearance modification.

LangSurf \cite{li2024langsurflanguageembeddedsurfacegaussians} addresses the misalignment of language features with 3D surfaces seen in methods like LeRF and LangSplat, which hampers 3D querying, segmentation, and editing. To improve feature quality and contextual richness, it introduces a Hierarchical-Context Awareness Module. This module uses a pretrained image encoder to extract pixel-aligned features and applies hierarchical mask pooling over small, medium, and large object masks from SAM. By averaging features within these masks, the system generates context-aware embeddings that preserve object-scale semantics and perform better in low-texture areas. These embeddings are compressed via an autoencoder to reduce memory and training cost. The pooled features guide the training of a language-embedded surface field, where each Gaussian is assigned a CLIP-aligned semantic vector. Training begins with RGB supervision, then progresses to semantic and geometric learning using spatial and grouping constraints, including instance-aware strategies to distinguish similar-category objects.

FMGS \cite{zuo2025fmgs} integrates CLIP and DINO into 3D Gaussian Splatting using multi-resolution hash encoding (MHE) for open-vocabulary scene understanding. Instead of embedding features per Gaussian, FMGS queries the MHE-based semantic field using Gaussian positions, decoding them via an MLP into compact semantic vectors drastically reducing memory usage. These vectors are trained using multi-scale CLIP features, with a pixel alignment loss using DINO to boost spatial coherence. At inference, cosine similarity between prompts and rendered CLIP features yields softmax-normalized relevancy scores, enabling accurate language-guided localization.

LEGS \cite{yu2024language} builds on this by introducing scale-aware language features for real-time, room-scale 3D scene understanding, targeting mobile robotics. Like FMGS, it avoids per-Gaussian embeddings and queries a hash-encoded semantic field via an MLP. Inspired by LERF and LERF-TOGO \cite{lerftogo2023}, LEGS supports multi-scale semantic reasoning by conditioning on both 3D position and object scale, enabling part-level understanding, faster training, and querying through the speed benefits of Gaussian Splatting.

FastLGS \cite{ji2025fastlgs} further enhances performance by replacing per-Gaussian embeddings with a grid-based mapping strategy. Multi-view CLIP features, extracted via SAM masks, are stored in a structured 3D semantic feature grid trained alongside 3DGS. During inference, these grids reconstruct pixel-level CLIP features, allowing real-time, zero-shot localization. FastLGS achieves dramatic speedups of up to 98× over LERF and 4× over LangSplat while maintaining comparable semantic accuracy, demonstrating the efficiency of grid-based language feature representation.

OCCAM \cite{cheng2024occam} introduces a training-free approach to language-guided 3D Gaussian Splatting by lifting 2D semantic features into 3D without optimization. Instead of learning per-Gaussian embeddings, the method treats semantic lifting as a maximum-likelihood estimation problem. Each Gaussian's 3D semantic feature is computed as a weighted average of its projected 2D CLIP features across multiple views, using $\alpha$-blending weights. Assuming dominant per-pixel Gaussian contributions and negligible cross-Gaussian interactions, OCCAM avoids costly optimization while supporting uncompressed 512D CLIP features guided by SAM-based multilevel masks.

GaussianGrasper \cite{zheng2024gaussiangrasper} proposes a contrastive feature distillation framework for training a 3D language field from RGB-D inputs. Each Gaussian is initialized with a learnable latent vector, decoded into CLIP space via an MLP. SAM-generated instance masks supervise these embeddings using a contrastive loss that enforces intra-mask consistency and a distillation loss aligning rendered features with CLIP. Unlike LangSplat, which uses a scene-specific autoencoder for feature compression, GaussianGrasper omits the encoder and jointly optimizes the latent vectors and decoder directly.

GOI \cite{qu2024goi} advances feature-efficient semantic reasoning by separating compression from filtering through two novel components: a Trainable Feature Clustering Codebook (TFCC) and an Optimizable Semantic-space Hyperplane (OSH). Instead of per-Gaussian CLIP features or learned latent vectors alone, each Gaussian holds a 10D latent code, decoded into logits to select a 256D embedding from the TFCC. This clustering mechanism promotes semantic sharing across Gaussians. Meanwhile, OSH learns a query-specific hyperplane in semantic space to filter relevant Gaussians in a differentiable, data-driven manner. GOI’s decoupled design contrasts with GaussianGrasper’s per-Gaussian optimization and LangSplat’s scene-level autoencoding, offering improved scalability and semantic precision.

Semantic Gaussians \cite{guo2024semantic} extend 3D Gaussian Splatting for open-vocabulary scene understanding by injecting a semantic component into each Gaussian. Semantic features from pre-trained VLMs (CLIP, OpenSeg, VLPart) are projected onto 3D Gaussians via a training-free pixel-to-Gaussian mapping based on spatial correspondence. To improve generalization and efficiency, a 3D semantic network based on MinkowskiNet is trained to predict these semantic components directly from raw Gaussians, supervised by the projected features. This design supports zero-shot generalization to unseen scenes and enables diverse tasks like part- and instance-segmentation, tracking, and scene editing. 

ChatSplat \cite{chen2024chatsplat} enables multi-level conversational interaction in 3D by augmenting Gaussians with view-level and object-level language features from LLaVA embeddings. A scene-specific autoencoder compresses and aligns these high-dimensional features with the 3D Gaussian space, aided by a scaling strategy for compatibility with LLM embeddings. During inference, language features are rendered into 2D maps via a tile-based rasterizer, then tokenized using a two-stage encoder. First, a scene-specific CNN for reduction, followed by a non-overlapping CNN for tokenization. Unlike LangSplat, which uses implicit pixel-wise embeddings, ChatSplat decouples object masks from feature maps, enabling targeted, context-aware object dialogue and shifting the focus from passive tasks to interactive language grounding in 3D.

SplatTalk \cite{thai2025splattalk} focuses on generalizable 3D scene understanding via a shared autoencoder trained across multiple scenes, supporting zero-shot VQA without explicit 3D supervision (e.g., depth or point clouds). Visual-language tokens from multi-view RGB images are extracted using LLaVA-OV and compressed into latent features embedded into Gaussians. These jointly optimize scene geometry and semantics. At inference, entropy-based sampling selects informative Gaussians, which are then projected back into the LLM’s input space for direct language querying. SplatTalk matches the performance of 3D-LLMs trained with geometric data, offering a scalable, geometry-free solution for language-guided 3D scene interaction.

OpenGaussian \cite{wu2024opengaussian} extends 3DGS to support point-level open-vocabulary understanding, addressing the limitations of prior methods that rely on 2D rendering for language supervision (e.g., LangSplat \cite{qin2024langsplat}, LangSurf \cite{li2024langsurflanguageembeddedsurfacegaussians}, ChatSplat \cite{chen2024chatsplat}). Instead of lifting 2D CLIP features via per-scene optimization, OpenGaussian directly learns instance-discriminative representations for each Gaussian using intra-mask smoothing and inter-mask contrastive losses, guided by SAM masks. A coarse-to-fine codebook discretizes these features by clustering Gaussians both spatially and semantically, promoting instance consistency. To avoid neural compression, it introduces a training-free 2D–3D association based on IoU and feature similarity, allowing the use of full-resolution CLIP embeddings. While effective for segmentation and object selection, OpenGaussian still requires per-scene training and codebook construction, limiting scalability.

Dr. Splat \cite{jun2025dr} developed concurrently, targets the same goal, but eliminates all scene-specific training. It introduces a training-free inverse feature registration strategy that directly maps CLIP embeddings to 3D Gaussians. Instead of learning instance features or per-scene codebooks, Dr. Splat projects CLIP features onto the top-k Gaussians intersected by each pixel ray, aggregating them across views via visibility-weighted fusion. These features are L2-normalized and compressed using Product Quantization (PQ) with a scene-agnostic codebook pre-trained on LVIS \cite{gupta2019lvis}, enabling compact storage without losing semantic richness. At inference, text queries are encoded with CLIP and matched to PQ-decoded Gaussian embeddings via cosine similarity and lightweight re-ranking. This fully training-free design supports efficient 3D querying, segmentation, and localization with strong performance, low memory cost, and high cross-scene generalization.

\subsection{Dynamic Scenes}

Recent advances in volumetric rendering and multimodal modeling have enabled spatiotemporal understanding of dynamic 3D scenes through natural language.

DGD \cite{labe2024dgd} introduces a unified 3D representation using deformable 3D Gaussians that capture both appearance and semantics from monocular video input. Semantic features from 2D foundation models (e.g., DINOv2, CLIP) are directly embedded into each Gaussian, enabling open-vocabulary interaction via text or clicks. Unlike methods that compress features, DGD stores full high-dimensional vectors (e.g., 384D/512D), simplifying architecture but increasing memory use. The model jointly optimizes spatial, appearance, and semantic parameters over time, supporting segmentation, tracking, and region-based editing.

4-LEGS \cite{fiebelman20254} extends Gaussian Splatting to 4D, grounding language in dynamic scenes. It uses ViCLIP to extract multiscale video-language features, which are compressed into a low-dimensional latent space via a scene-specific autoencoder. These latent embeddings are assigned to Gaussians at each timestep, enabling dynamic 3D representation with embedded semantics. At inference, text queries are matched against the 4D field using a relevance score between encoded text and decoded latent features, supporting tasks like scene editing and semantic video search. 

4D LangSplat \cite{li20254d} targets both static and dynamic language queries by building two semantic fields: one for object categories (e.g., “cup”) and one for temporal actions (e.g., “sitting down”). Instead of visual embeddings, it uses MLLMs to generate object-level captions based on visual and textual prompts. These captions are embedded and used as supervision for per-object, pixel-aligned features. A scene-specific autoencoder compresses these features, while a status deformable network models each Gaussian’s semantic state over time as a blend of learned prototypes. This enables precise, open-vocabulary querying across time-varying 3D scenes.

\section{Real World Applications}

\begin{figure*}[!htbp]
    \centering    \includegraphics[width=0.75\textwidth]{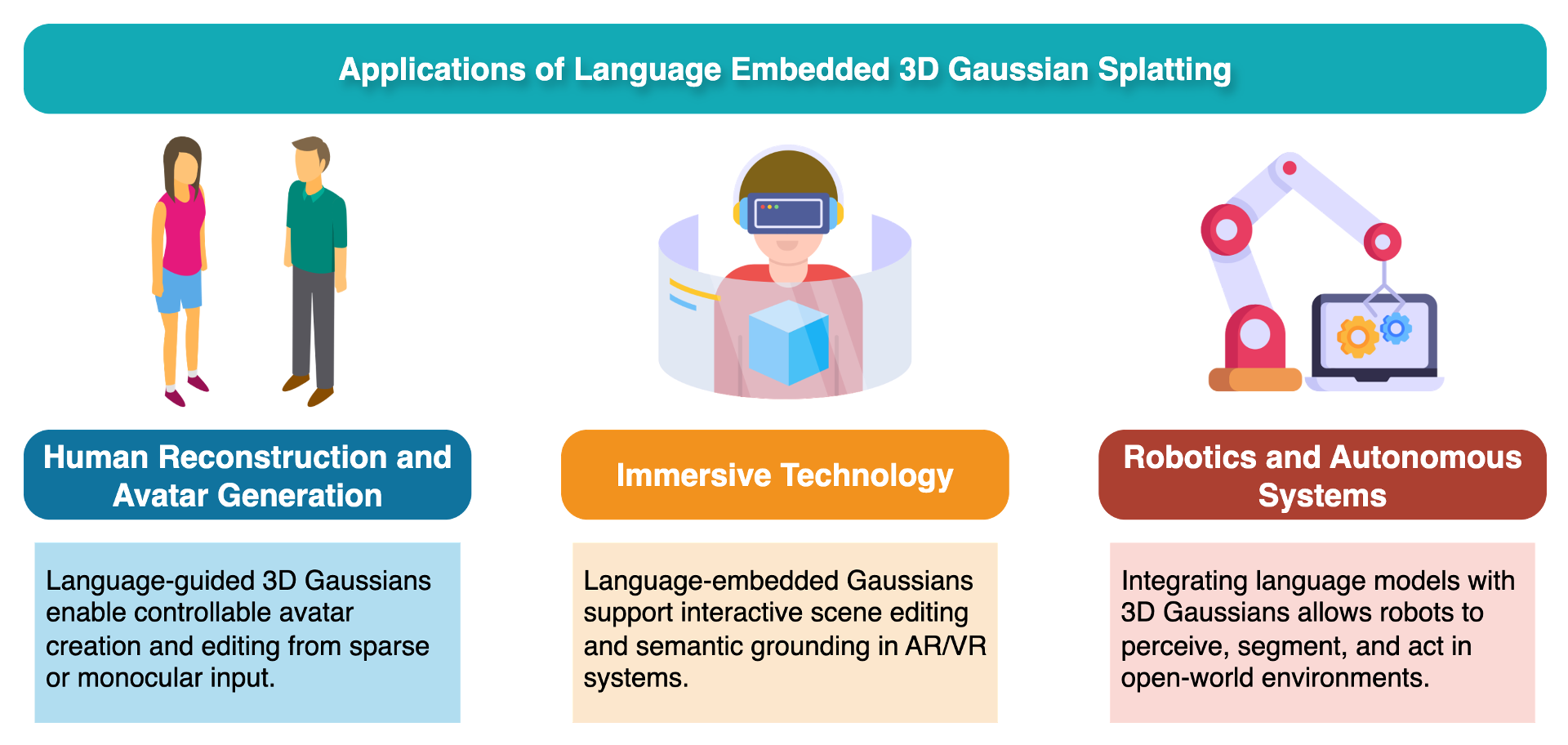}
\caption{Application areas of language-embedded 3D Gaussian Splatting across humans, robots, and immersive systems.}
\Description{Application areas of language-embedded 3D Gaussian Splatting across humans, robots, and immersive systems.}
\label{fig:applications}
\end{figure*}

While the previous sections explored how 3DGS can be integrated with language and foundation models to support tasks such as semantic supervision, open-vocabulary querying, and interactive scene generation, these capabilities are no longer limited to controlled or experimental settings. Recent work demonstrates the growing impact of language-embedded 3DGS across a wide range of real-world applications. From avatar generation and immersive virtual environments to robotics and autonomous systems, these methods enable systems to perceive, generate, and interact with complex 3D environments in more intelligent and context-aware ways. Fig. \ref{fig:applications} depicts these application domains, highlighting how multimodal integration with 3DGS extends its functionality beyond rendering to support high-level reasoning and interaction in both virtual and physical settings.

\subsection{Human Reconstruction and Avatar Generation}

Modeling humans in 3D is a vital task across multiple industries, including entertainment, gaming, fashion, and e-commerce. Whether for digital avatars in virtual worlds, realistic virtual try-on systems, or expressive animated characters, accurately capturing the complexity of human appearance, motion, and expression remains a significant challenge. While mesh-based and parametric models (such as SMPL \cite{loper2023smpl, pavlakos2019expressive} and imGHUM \cite{alldieck2021imghum}) provide a structured representation of body shape and pose, they often fail to capture fine-grained details such as facial expressions, skin texture, or clothing deformation \cite{yang2024innovative}. Recent works address these limitations by combining 3DGS with parametric priors and language-based editing, enabling high-quality, customizable human avatars \cite{liu2024humangaussian, moon2024expressive, gong2024laga, kocabas2024hugs}.

Several methods incorporate diffusion-based optimization guided by text prompts to generate 3D avatars with realistic appearance and geometry \cite{liu2024humangaussian, gong2024laga, yuan2024gavatar}. Others focus on controllable head avatars, stylization, or emotion-driven animation using vision-language models such as CLIP or BLIP-2 \cite{zhou2024headstudio, liu2024efficient, jiang2024fast}. For applications like virtual try-on, hybrid pipelines leverage latent diffusion, ControlNet, or LoRA-based personalization to support garment-specific edits while maintaining multi-view consistency \cite{chen2024gaussianvton, cao2024gs}.

Together, these systems demonstrate the value of combining 3DGS with foundation models to enable expressive, dynamic, and user-controllable human reconstruction.

\textbf{2D-to-3D Generation:}
An emerging trend in human reconstruction involves using 2D generation models to synthesize multiple views of a subject, which are then integrated into 3D reconstruction pipelines \cite{cao2024gs, chen2024gaussianvton, jiang2024fast}. Diffusion models, such as Stable Diffusion \cite{rombach2022high} and DALL-E \cite{ramesh2021zero}, have shown remarkable success in generating high-quality images from textual descriptions, enabling the generation of various 2D perspectives. These 2D images, when combined with NeRFs or other view synthesis techniques, can be used to create a detailed 3D model. Gaussian splatting plays a critical role in refining these 3D models by providing smooth surface representations, helping to eliminate artifacts that often arise from 3D rendering techniques that rely solely on meshes or voxel grids. 

\textbf{Integration of LLMs and Foundation Models:}
An emerging research area focuses on integrating LLMs and Foundation models with 3D reconstruction pipelines to enable text-based avatar generation. LLMs can interpret complex natural language descriptions and translate them into 3D attributes, such as body shape, clothing, and facial expressions. When combined with VLMs, such as CLIP, which align textual descriptions with visual content, this approach enables users to customize avatars through simple language commands. These avatars can then be generated in real time, with 3DGS ensuring smooth and realistic rendering of the human model. This integration opens the door for more intuitive, accessible avatar creation, where users no longer need to interact with complex modeling software. Relevant studies illustrating these approaches are summarized in Table \ref{tab:avatar-summary}.

The integration of 3DGS, human priors, and 2D-to-3D generation methods represents a promising direction for advancing human reconstruction and avatar generation. By combining the parametric precision of models like SMPL with the fluid rendering capabilities of Gaussian splatting, researchers can create highly realistic, customizable avatars that can be easily manipulated and animated. Furthermore, the incorporation of LLMs and VLMs for text-based input allows for a seamless user experience, offering unparalleled flexibility in avatar creation. Continued advancements in these areas will enable more immersive, dynamic virtual experiences and open new possibilities for real-time, personalized avatars in gaming, entertainment, and beyond.

\begin{table*}[htbp]
\centering
\tiny
\renewcommand{\arraystretch}{1.1}
\setlength{\tabcolsep}{5pt}
\begin{tabular}{p{0.4cm} p{1.5cm} p{1.8cm} p{6.2cm} c c}
\toprule
\textbf{Study} & \textbf{Application} & \textbf{AI/Language Model} & \textbf{Key Contributions} & \textbf{Open Voc.} & \textbf{Dyn. Scene} \\
\midrule
\cite{liu2024humangaussian} & Human/Avatar Generation & Stable Diffusion (SDS) & Structure-Aware SDS combining RGB/depth; Annealed negative prompt guidance; SMPL initialization with adaptive pruning & \checkmark & \\
\cite{gong2024laga} & Human/Avatar Generation & Stable Diffusion (dual-SDS) & Coarse-to-fine generation; Layered Gaussian garments; Garment transfer via human fitting, similarity, and visibility loss & \checkmark & \\
\cite{yuan2024gavatar} & Human/Avatar Generation & SDS & Primitive-based Gaussian animation; Implicit attribute fields; SDF-driven mesh extraction; Real-time rendering (100 fps) & \checkmark & \checkmark \\
\cite{jiang2024fast} & Head Generation & Diffusion, CLIP & Style-aligned sampling loss; CLIP-based stylization control; Single image-driven 3DGS with stylized portrait rendering & \checkmark & \\
\cite{liu2024efficient} & Head Generation & CLIP & Emotion-conditioned deformation fields; Dynamic rendering from audio/emotion; Temporal fusion of 3DGS features & \checkmark & \checkmark \\
\cite{zhou2024headstudio} & Head Generation & Realistic Vision 5.1, ControlNet, MediaPipe, SDS & FLAME-driven Gaussian animation; Dense initialization; Distillation via SDS and landmark priors; Real-time dynamic head avatars & \checkmark & \checkmark \\
\cite{chen2024gaussianvton} & Virtual Try-On & LaDI-VTON (Latent Diffusion) & Three-stage editing with ControlNet; Edit Recall Reconstruction (ERR); Garment-coherent refinement with multiview alignment & & \\
\cite{cao2024gs} & Virtual Try-On & Stable Diffusion (Inpainting), SDS, LoRA, BLIP-2 & Persona-consistent 3D editing; LoRA-driven reference matching; First 3D VTON benchmark; Garment style/pose control from prompts & & \\
\bottomrule
\end{tabular}
\caption{Comparison of human/avatar generation papers combining language and Gaussian Splatting by application, foundation model usage, technical contributions, and support for open-vocabulary queries and dynamic environments.}
\label{tab:avatar-summary}
\end{table*}
\subsection{Immersive Technology}
Immersive technologies such as Virtual Reality (VR), Augmented Reality (AR), and Mixed Reality (MR), collectively referred to as Extended Reality (XR), enable rich, real-time interaction with 3D environments \cite{adriana2022}. When combined with language-embedded 3DGS, these systems support high-fidelity reconstruction, semantic understanding, and intuitive user interaction across a range of XR applications.

\textbf{3DGS in Extended Reality:}
3DGS offers real-time, high-fidelity scene representation, making it a strong candidate for XR applications. Its compact and explicit representation supports photorealistic rendering, contributing to immersive virtual environments. However, challenges remain around physical interaction, real-time performance, and integration with XR development tools.

\citet{qiu2025creatingvirtualenvironments3d} demonstrated a 3DGS-based pipeline for VR, achieving high visual quality but lacking physics-based interactions. \citet{vr_gs_2024} addressed this by incorporating segmentation, inpainting, mesh generation, and physical attributes via PhysGaussian \cite{xie2024physgaussian}, enabling dynamic object interactions, although at high computational cost.

To improve efficiency, recent work has focused on optimizing rendering and memory usage. \citet{tu2024} reduced temporal artifacts and improved frame rates, while \citet{kim2024} used superpixel-guided sampling to reduce Gaussian counts, lowering GPU memory consumption by 2–3×. \citet{iandola2024squeezeme} compressed full-body Gaussian avatars for real-time animation at 72 FPS on mobile VR (Meta Quest 3), using a Vulkan-based rendering pipeline. 

In AR, 3DGS is used for localization and avatar rendering. \citet{zhai2024splatloc3dgaussiansplattingbased} introduced SplatLoc, aligning live camera feeds with precomputed Gaussian models for real-time pose estimation. \citet{chen2025taoavatarrealtimelifelikefullbody} used 3DGS to render lifelike avatars on the Apple Vision Pro, integrating a teacher-student model for real-time performance and language-driven interaction for mixed reality experiences.

While integration with physics engines and XR SDKs remains an open challenge, ongoing work, including hybrid mesh-splat models \cite{Dongye2025}, suggests promising pathways toward more interactive and physically grounded XR experiences.

\textbf{Integration of LLMs and Foundation Models:}
LLMs have rapidly transformed numerous industries by enabling machines to understand and generate human language with unprecedented fluency, and Extended Reality is no exception. In XR environments, LLMs (and other language models) are increasingly being used for a wide range of applications such as creating conversational characters \cite{maslych2025takeawaysapplyingllmcapabilities, yousri2024illusionxllmpoweredmixedreality}, enabling dynamic storytelling \cite{constantinides2024}, powering adaptive learning experiences \cite{yousri2024illusionxllmpoweredmixedreality}, supporting intuitive object manipulation \cite{wang2025therellmbasedvrmover}, and facilitating human-robot interactions \cite{zhang2025llmdrivenaugmentedrealitypuppeteer}. The integration of 3DGS and language offers promising advancements towards more intelligent and responsive virtual environments, demonstrated by the recent studies summarized in Table \ref{tab:xr-summary}.

\begin{table*}[htbp]
\centering
\tiny
\renewcommand{\arraystretch}{1.1}
\setlength{\tabcolsep}{5pt}
\begin{tabular}{p{0.4cm} p{1.7cm} p{1.9cm} p{6.2cm} c c}
\toprule
\textbf{Study} & \textbf{Application} & \textbf{AI / Language Model} & \textbf{Key Contributions} & \textbf{Open Voc.} & \textbf{Dyn. Scene} \\
\midrule

\cite{gu2025vrsketch2gaussian} & Sketch-Guided 3D Generation & CLIP & Two-stage sketch-text alignment framework for 3D object generation using 3DGS; Geometry control via VR sketches and appearance control via text prompts; Release of VRSS dataset & \checkmark &  \\

\cite{mao2024livegsllmpowersinteractive} & Interactive Physics Simulation & GPT-4o, SAM, DEVA & LIVE-GS enables real-time physical interaction in 3DGS-based VR environments using GPT-4o for material analysis; introduces a feature-mask segmentation strategy; proposes a unified PBD-based simulation framework supporting rigid, soft, and granular materials. & \checkmark & \checkmark \\

\cite{Guo2024} & Sensor-Driven AR Scene Generation & GPT-3.5, GPT-4 & LLM-driven framework that interprets IoT sensor data to generate textual scene descriptions; synthesizes corresponding AR visualizations using text-to-3D tools; proposes a benchmark for evaluating scene descriptiveness and AR coherence. & \checkmark &  \\

\cite{bowser2024d} & Telepresence for Human-Robot Interaction & Language Prompts (model unspecified), SAM & 3DGS-based HRI interface for disaster scenarios; Semantic overlays using SAM in the reconstructed environments; supports real-time exo- and ego-centric views and natural language querying of the robot's environment. & \checkmark & \checkmark \\

\bottomrule
\end{tabular}
\caption{Comparison of immersive technology applications combining language models and Gaussian Splatting, categorized by use case, AI model integration, and support for open-vocabulary querying or dynamic scene interaction.}
\label{tab:xr-summary}
\end{table*}

One notable example of combining language and 3DGS within immersive technologies is VRSketch2Gaussian \cite{gu2025vrsketch2gaussian}, a framework that enables 3D object generation from multimodal inputs in virtual reality. The system allows users to provide both a freehand VR sketch and a textual prompt, fusing geometric and semantic input. While the VR sketch serves as the primary geometric prior, textual descriptions play a crucial role in refining abstract attributes such as color, texture, and material. The architecture integrates these modalities by extracting text embeddings via CLIP and concatenating them with the VR sketch embeddings. This fused representation is then passed through a cross-attention module within a 3D U-Net, which acts as the denoising network in the 3D diffusion model responsible for generating the final set of Gaussian splats. This approach exemplifies how natural language can enrich immersive 3D content creation by guiding not just structure but also stylistic and semantic detail.

Another use case is introduced by \citet{mao2024livegsllmpowersinteractive} with LIVE-GS, a framework for 3DGS scene representation that integrates GPT-4o to infer object properties directly from images, eliminating the need for manual tuning required by previous methods such as \cite{vr_gs_2024}. Built on the Position-Based Dynamics (PBD) method \cite{Muller2007}, LIVE-GS leverages LLMs for material analysis, artifact tracking, and scene inpainting. Designed and tested specifically for virtual reality, the system enables immersive, real-time interactions within reconstructed environments by combining efficient physical simulation with GPT-4o's multimodal reasoning capabilities.

Building on the potential of LLMs and 3DGS in dynamic XR environments, \citet{Guo2024} introduced Sensor2Scene, a framework that integrates LLMs with 3DGS-based generation tools. By combining multimodal sensor data with LLMs, the system generates detailed, context-aware scene descriptions that enhance AR situational awareness and enable adaptive interactions. Sensor2Scene features an AI-driven adaptive visualization engine that dynamically updates rendered content in response to real-time sensor inputs, environmental changes, and user preferences. For 3D content generation, the framework uses models like DreamGaussian \cite{tang2023dreamgaussian} to generate responsive and visually realistic elements within AR environments.

In the area of human-robot interaction (HRI) \citet{bowser2024d} explore how Gaussian Splatting and language prompts can improve virtual reality based disaster response training and mobile robot navigation. Their system provides both ego- and exo-centric views for the operator within a 3DGS-rendered virtual environment. Additionally, it integrates SAM for semantic scene segmentation, allowing operators to identify objects of interest. Meanwhile, NLP integration makes the environment ``smart,'' enabling the efficient identification of specific locations and hazardous objects in realistic large-scale environments. 

\subsection{Robotics and Autonomous Systems}

Language-embedded 3DGS frameworks have introduced new possibilities in robotics, where systems must interpret complex environments, adapt to changes in real time, and reason over high-level task descriptions. Traditional approaches have relied on modular perception-control pipelines with limited semantic flexibility. By embedding both geometric and semantic representations in the same data structure, 3DGS-based methods enhanced with foundation models offer a more unified approach to robotic perception and interaction. This section reviews recent works that explore this intersection across SLAM, robotic manipulation, and language-conditioned control tasks. Table \ref{tab:robotics-summary} summarizes the related studies, highlighting model integration, system capabilities, and contributions.

\begin{table*}[htbp]
\centering
\tiny
\renewcommand{\arraystretch}{1.1}
\setlength{\tabcolsep}{5pt}
\begin{tabular}{p{0.4cm} p{1.5cm} p{1.5cm} p{6.5cm} c c}
\toprule
\textbf{Study} & \textbf{Application} & \textbf{AI/Language Model} & \textbf{Key Contributions} & \textbf{Open Voc.} & \textbf{Dyn. Scene} \\
\midrule
\cite{zheng2024gaussiangrasper} & Robotic Grasping & CLIP, SAM & Efficient Feature Distillation (EFC) using contrastive learning; Feature field creation from SAM; Normal-guided grasp filtering; Scene update mechanism & \checkmark & \\
\cite{ji2024neds} & Semantic SLAM & DINO, DepthAnything & Semantic-depth fusion for consistent 3D features; High-dim compression into Gaussians; Virtual camera view pruning; Joint multi-channel supervision across modalities & & \\
\cite{yu2024language} & Object Localization \& Mapping & CLIP & Online multi-camera mapping; Multi-resolution hashgrid encoding; Incremental bundle adjustment; CLIP-based semantic embedding with geometric fusion & \checkmark & \\
\cite{lu2024manigaussian} & Robotic Manipulation & Stable Diffusion, PerceiverIO & Temporal-aware Gaussian splats for future state prediction; Multimodal transformer for language-conditioned action prediction; Multi-objective training loss & \checkmark & \checkmark \\
\cite{xu2024gaussianproperty} & Grasping \& Simulation & SAM, GPT-4V & SAM-based part-level segmentation; GPT-4V for physical property reasoning; Multi-view 2D to 3D voting strategy for physical annotation; Safe grasp force prediction module & & \\
\cite{yu2025persistent} & Robotic Manipulation & Detic, DINOv2, CLIP & Combines Detic/CLIP/DINOv2 for feature-rich Gaussians; Supports segmentation, 6-DoF pose tracking, and language querying; Fast online updates & \checkmark & \checkmark \\
\cite{lan2024monocular} & Monocular SLAM & CLIP & Sliding window multi-view optimization; CLIP-based loop closure for drift correction; Text-to-trajectory localization; Graph backend for global consistency & \checkmark & \\
\cite{ji2024graspsplats} & Robotic Manipulation & MobileSAM, CLIP & Hierarchical reference features with MobileSAM and MaskCLIP; Efficient language-guided grasping via object/part-level queries; Real-time deformable splats with Kabsch alignment & \checkmark & \checkmark \\
\cite{pham2024go} & Semantic SLAM & ChatGPT-4o, DINO, SAM & ChatGPT-4o for label generation; DINO for detection and SAM for segmentation; PRM-based path planning; Back-projection based 3D object localization & \checkmark & \\
\cite{shorinwa2024splat} & Robotic Manipulation & CLIP, VRB & ASK-Splat for embedding semantics and affordances; SEE-Splat for editable scene rendering and real-time updates; Grasp-Splat for affordance-aligned grasping using GraspNet & \checkmark & \checkmark \\
\cite{sheng2024msgfield} & Robotic Manipulation & CLIP, SAM, YOLO-World & Motion basis decomposition with shared semantic embeddings; Occlusion-aware grasp filtering; Dynamic re-optimization with 2DGS representation & \checkmark & \checkmark \\
\cite{chahe2025query3d} & Autonomous Driving & GPT-3.5, Qwen2.5, LLaMA 3.2 & LLM-generated canonical and helping positive words for better context; Fine-tuned small LLMs for faster on-device inference; Enhanced segmentation via LE3DGS integration & \checkmark & \\
\cite{chen2025splat} & Navigation & CLIP, Lang-Splat & Splat-Plan planner with safe polytope corridor generation; Splat-Loc for RGB-only pose estimation via PnP; Real-time closed-loop re-planning with language goals & \checkmark & \\
\cite{li2024hier} & Semantic SLAM & GPT-4o-mini & LLM-guided hierarchical semantic tree coding; Hierarchical loss for scalable category inference; Efficient large-scale scene modeling & & \\
\bottomrule
\end{tabular}
\caption{Comparison of robotics-related papers combining language and Gaussian Splatting by application domain, model usage, technical contributions, and support for open-vocabulary queries and dynamic environments.}
\label{tab:robotics-summary}
\end{table*}

\textbf{Scene Understanding and SLAM:}
Several studies employ GS to augment SLAM with semantic awareness and language grounding. LEGS \cite{yu2024language} incrementally reconstructs indoor environments using a multi-camera setup on a mobile robot, embedding CLIP-derived features into Gaussians for open-vocabulary object localization. Monocular Gaussian SLAM \cite{lan2024monocular} focuses on improving loop closure and global consistency using CLIP, enabling trajectory relocalization from language prompts such as "return to the hallway with the blue chair." Go-SLAM \cite{pham2024go} combines ChatGPT-4o with Grounding DINO and SAM to identify and segment objects from natural language queries and integrate them into a navigable 3DGS-based map. This system also incorporates probabilistic-based motion planning, making the outputs actionable for autonomous navigation. Hier-SLAM \cite{li2024hier} contributes a hierarchical semantic labeling method using LLM-generated tree codes. Although it does not directly support user queries, it scales well to large and diverse environments. In contrast, NEDS-SLAM \cite{ji2024neds} emphasizes dense semantic mapping by fusing DINO features with DepthAnything to support spatially consistent training without requiring language input. Together, these works demonstrate the feasibility of embedding semantic reasoning and symbolic reference into GS-based spatial memory.

\textbf{Robotic Manipulation and Grasping:}
In robotic manipulation, 3DGS has been used to represent objects and interaction contexts with rich semantic attributes. GaussianGrasper \cite{zheng2024gaussiangrasper} introduces a feature distillation framework that uses CLIP and SAM to construct feature fields aligned with object geometry, supporting open-vocabulary grasping and normal-guided filtering. GraspSplats \cite{ji2024graspsplats} builds on this by incorporating part-level language queries, deformable object tracking, and hierarchical CLIP-based descriptors, enabling real-time grasping of moving or articulated objects. POGS \cite{yu2025persistent} fuses Detic, DINOv2, and CLIP to produce persistent, language-aware Gaussians capable of tracking object pose during manipulation. MSGField \cite{sheng2024msgfield} introduces a 2D Gaussian splatting representation that models motion, semantics, and geometry jointly. Through motion basis decomposition and occlusion-aware filtering, it supports interaction in dynamic environments. GaussianProperty \cite{xu2024gaussianproperty} brings in physical reasoning by combining GPT-4V and SAM to infer material properties such as density and friction at the part level, which are then projected into the 3DGS structure for use in force-aware grasp planning and simulation. These systems collectively shift the manipulation paradigm toward task and context-aware representations, integrating symbolic understanding with geometric affordances.

\textbf{Language-Guided Planning and Multi-Stage Control:}
Beyond scene understanding and manipulation, several recent works further explore the potential of language and foundation models to drive multi-stage or predictive robotic behavior. Splat-MOVER \cite{shorinwa2024splat} combines language-guided scene understanding, interactive editing, and grasp planning in a unified 3DGS pipeline. The system integrates CLIP and VRB \cite{bahl2023affordances} features across three modules (ASK-Splat, SEE-Splat, and Grasp-Splat) to support commands such as "move the blue mug to the upper shelf and place it upright." ManiGaussian \cite{lu2024manigaussian} takes a predictive approach, learning temporal representations of object behavior from multimodal data and Stable Diffusion supervision. The model uses a transformer architecture to anticipate scene changes and produce language-conditioned action plans based on predicted dynamics. Splat-Nav \cite{chen2025splat} applies 3DGS to low-altitude drone navigation. It introduces Splat-Plan, a path planner that leverages the ellipsoidal structure of GS to construct safe corridors, and Splat-Loc, a lightweight pose estimator using only RGB input. The system supports text-specified goals while maintaining real-time localization and replanning capabilities.

These robotics-oriented studies reveal several converging trends. Most systems incorporate vision-language models such as CLIP or DINO to map visual inputs to open-vocabulary semantics, enabling flexible task specification through natural language. Many approaches now support inference-time text queries or commands, reducing the need for task-specific retraining and enhancing generalization across environments. A subset of systems, including GraspSplats, MSGField, Splat-MOVER, and ManiGaussian, explicitly address the challenges of dynamic scene understanding and real-time adaptation. Across applications, the most capable systems tightly couple geometry, semantics, and temporal reasoning within a unified Gaussian Splatting framework. This integration reflects a broader shift toward representations that support both symbolic intent interpretation and low-level control execution.

By embedding structure and semantics into a common representation, 3DGS-based systems augmented with foundation models offer a flexible interface between high-level language input and grounded robotic action. Although current implementations are often specialized to particular tasks or environments, the architectural patterns emerging from this research point toward more general-purpose, language-driven autonomy. Future developments may further align scene understanding, planning, and control through integrated representations that support multimodal grounding, dynamic world modeling, and adaptive decision making.

\section{Challenges and Future Research Directions}
Despite recent progress in language-embedded 3DGS, several fundamental challenges remain. These span architectural, computational, and dataset-related limitations, as well as unresolved issues in semantic fidelity, spatial reasoning, and generalization. In this section, we highlight key open research problems and outline promising directions to address them, with a focus on improving scalability, robustness, and semantic alignment across real-world scenarios.
\subsection{Photorealism vs. Semantic Fidelity in 3DGS}

While 3DGS delivers state-of-the-art photorealistic rendering with high efficiency, it remains prone to artifacts known as floaters, Gaussian primitives not anchored to actual surfaces. These typically arise from sparse reconstructions, inaccurate geometry, or optimization instability during training. Although floaters may be visually negligible in some contexts, they pose serious challenges for multimodal and semantic tasks.

In particular, floaters introduce high-frequency noise and semantic ambiguity, which can degrade performance in applications involving feature distillation from vision-language or foundation models. They may attract unintended attention or distort spatial correspondence, undermining tasks like grounding, captioning, and attention pooling. As 3DGS gains traction in 3D scene understanding and vision-language reasoning, addressing floaters becomes essential for ensuring reliable semantic inference.

\textbf{Future Research Direction:}  To improve semantic fidelity, future research should explore geometry-aware denoising methods that reduce or eliminate floaters while preserving critical visual and semantic information. Approaches such as filtering based on surface proximity, visibility-based pruning, confidence-weighted density scaling, or learned occupancy priors show promise for aligning semantic features with geometrically valid regions. Such methods enhance the robustness and interpretability of 3DGS in language-driven and embodied AI applications.

Recent studies, including StableGS \cite{wang2025stablegs}, Pixel-GS \cite{zhang2024pixel}, and FreeSplat++ \cite{wang2025freesplat++}, have begun tackling these challenges by introducing effective floater suppression techniques. Incorporating these advances into multimodal pipelines can significantly boost the semantic reliability of 3DGS-based representations in dynamic and complex environments.

\subsection{Vision-Language Models in 3D}

While VLMs excel in 2D tasks, their extension to 3D scene understanding remains fundamentally limited, not due to the choice of 3D representation (e.g., meshes, voxels, or 3D Gaussian Splatting), but because of how these models are trained. Most approaches project 3D scenes into 2D views and supervise them with features or captions from pre-trained 2D VLMs, inheriting their limitations, especially in spatial reasoning \cite{anis2025limitations}.

VLMs lack explicit understanding of spatial relationships such as "behind," "inside," or "to the left of," which are essential in 3D environments. Their reliance on language priors and co-occurrence statistics can result in semantically plausible but spatially incorrect predictions. Additionally, they tend to focus on global features, missing the fine-grained details crucial for 3D tasks that require dense correspondence across views \cite{chen2024spatialvlm, qi2025gpt4scene}.

A further limitation is the absence of geometric supervision. Trained solely on 2D data, VLMs lack awareness of depth, occlusion, or topology, making them ill-suited for tasks like scene reconstruction, navigation, or object interaction. While fine-tuning on 3D datasets can improve task-specific performance, it often reduces generalization, harming zero-shot capabilities; an issue for domains like robotics and embodied AI \cite{ha2022semantic}.

These challenges are most apparent in tasks like long-tail object localization or spatially grounded reasoning, where text prompts alone are insufficient. Even strong VLMs underperform in these settings without dedicated 3D spatial reasoning mechanisms \cite{ha2022semantic}.

\textbf{Future Research Direction}:
Advancing VLMs for 3D understanding will require training and architectural innovations that integrate semantic reasoning with spatial context. Instead of relying solely on rendered 2D views, future models should incorporate spatially aligned multimodal supervision, jointly embedding 2D, 3D, and language features. Representations like 3D Gaussian Splatting, with their differentiability, compactness, and multi-view consistency, offer a promising foundation.

To address the lack of depth and topology awareness, future models should include explicit geometric supervision such as occupancy fields, depth maps, and surface normals during training. These cues will help the model learn geometry-aware embeddings, improving occlusion handling and multi-view consistency for robust semantic understanding.

Rather than directly fine-tuning VLMs, which compromises generality, research should explore modular transfer learning approaches. Techniques like using frozen VLMs with spatial adapters or prompt tuning modules can inject 3D reasoning capabilities while preserving zero-shot performance. Recent efforts like SpatialVLM \cite{chen2024spatialvlm}, Semantic  Abstraction \cite{ha2022semantic}, and CG3D \cite{hegde2023clip} support this direction and could be extended to work with 3D Gaussian Splatting.

\subsection{Compute and Memory Bottleneck}
Language-guided 3DGS methods face substantial computational and memory challenges. Most pipelines require resource-heavy preprocessing steps, such as Structure-from-Motion (SfM) and COLMAP, followed by semantic supervision using large-scale VLMs and VFMs. These steps increase both training complexity and system overhead.

A major bottleneck stems from storing and processing high-dimensional language embeddings for every Gaussian point and camera view. As scenes can contain millions of Gaussians, each with attached feature vectors, the memory footprint grows rapidly, especially in large-scale or outdoor scenes. Many methods further perform feature field distillation, training the 3D representation to reproduce semantic features across multiple views, compounding compute and memory demands. These issues significantly limit training and inference on consumer-grade GPUs.

\textbf{Future Reserach Direction}:
To address these limitations, future work should aim to replace traditional SfM-based initialization with efficient, differentiable alternatives. SfM and COLMAP are non-differentiable, error-prone in complex environments, and introduce considerable preprocessing overhead making them ill-suited for scalable, language-guided 3DGS pipelines.

Emerging methods such as MAST3R \cite{leroy2024grounding}, DUST3R \cite{wang2024dust3r}, and VGGT \cite{wang2025vggt} provide promising end-to-end solutions, estimating geometry and camera poses without external tools. Integrating such approaches could enable fully differentiable workflows, reducing training time, improving robustness, and better aligning with VLM-based supervision. Progress in this direction would pave the way for scalable, semantically coherent 3DGS systems that operate efficiently in real-world conditions without the heavy computational burden of current methods.

\subsection{Generalizability:}
Current language-guided Gaussian Splatting approaches for 3D scene understanding typically operate in a scene-specific manner. These methods reconstruct a single scene and embed language features through feature field distillation, relying on supervision from large VLMs and VFMs. However, the learned semantic representations are not transferable across scenes. When the environment changes due to new objects, lighting, or layout, the entire pipeline must be retrained from scratch, limiting practical deployment in dynamic or large-scale settings.

\textbf{Future Reserach Direction}: To improve generalization, future work should explore advanced methods that learn semantics in a scene-agnostic manner without degrading the quality of 3D reconstruction rather than naively distilling 2D language embeddings directly into 3D Gaussian points. While techniques like neural semantic fields have shown promise in the NeRF domain, integrating similar approaches into the 3DGS pipeline could enhance both semantic expressiveness and generalization. Recent methods such as SceneSplat \cite{li2025scenesplat} and SceneSplat++ \cite{ma2025scenesplat++} aim to address this challenge directly. Instead of relying on per-point feature distillation, these methods apply scene-agnostic supervision through rendered views, using frozen VLMs. Their use of large-scale datasets and benchmarks demonstrates improved zero-shot performance across diverse scenes. However, further research is needed to develop language-guided 3DGS pipelines that are both semantically rich and robust across varied, real-world environments.

\subsection{Real-time Visualization Limitations:}
Most existing methods for querying semantic feature fields are limited to 2D image-space after rendering, with limited support for true 3D semantic interaction. While some approaches embed vision-language features into 3D Gaussians or lift 2D features into the 3D domain, these representations are often high-dimensional, abstract, and lack interpretable or interactive visualization tools. For example, methods like LERF, have demonstrated language-driven querying of 3D scenes through dedicated visualizers (e.g., built on top of NeRFStudio). While these tools may suffer from slow interaction on consumer-grade GPUs, they provide a foundation for semantic exploration in 3D. In contrast, equivalent real-time or interactive visualization tools are largely absent for language-guided 3DGS methods, limiting their usability in practical settings such as robotics, AR/VR, or embodied AI.  

\textbf{Future Research Direction}: While some efforts have been made to develop real-time visualizers such as in Feature3DGS \cite{zhou2024feature}, which uses the SIBR viewer, the current tools remain limited. Feature3DGS, for example, can only display all feature fields holistically and suffers from extremely slow performance on consumer-grade GPUs. It also lacks support for language-based interaction, such as prompting or grounding objects within a scene. Future work should focus on building more responsive, language-aware visualizers for 3DGS that enable real-time interaction and querying. This is essential not only for improving 3D scene understanding and visual grounding, but also for enabling downstream applications like language-guided scene editing, navigation, and embodied interaction.

\subsection{Datasets and Benchmarks}
Current language-guided 3D Gaussian Splatting methods are constrained by limited datasets and benchmarks. Most existing work relies on indoor datasets like ScanNet \cite{dai2017scannet} or Replica \cite{replica19arxiv}, which lack the diversity of real-world environments, including outdoor scenes with dynamic objects, varied lighting, and complex semantics. While outdoor datasets such as nuGrounding \cite{li2025nugrounding}, WildRefer \cite{lin2024wildrefer}, and NuScenes-QA \cite{qian2024nuscenes} exist, they are often domain-specific and provide minimal language supervision. In aerial contexts, datasets like UrbanScene3D \cite{UrbanScene3D} and DRAGON \cite{ham2024dragon} offer strong geometric data but lack the semantic annotations required for language-driven tasks.
\textbf{Future Research Direction}: Advancing language-guided 3DGS requires large-scale, semantically annotated 3D datasets that go beyond constrained indoor scenes. Extending existing datasets such as UrbanScene3D and DRAGON with high-quality language annotations could support tasks like grounding, captioning, and VQA. Additionally, standardized benchmarks that evaluate both geometric quality and semantic alignment would help track progress toward more generalizable, language-aware 3DGS pipelines.

\section{Conclusions}
In this paper, we surveyed the emerging intersection of 3D Gaussian Splatting and language-guided scene understanding. We began by outlining the 3DGS pipeline and summarizing advances in LLMs, VLMs, and VFMs. We then explored how these models, through their world knowledge, reasoning abilities, and zero-shot capabilities, can help address long standing challenges in 3D perception by enriching spatial understanding.

Building on this foundation, we reviewed recent methods that integrate LLMs, VLMs and VFMs into the 3DGS pipeline to enhance semantic reasoning and scene interpretation. We also discussed how 3D Gaussian Splatting has rapidly evolved from a rendering technique to a foundational component in broader applications, including 3D modeling, immersive environments, robotics, and autonomous systems.

Despite these advancements, the field still faces key limitations like high computation and memory demands, fragile semantic fidelity, limited generalization across scenes, and a lack of real-time interaction tools. These are compounded by the absence of large-scale, semantically annotated datasets for open-world 3D scene understanding. To address these challenges, we highlighted several future research directions aimed at building more efficient, scalable, and semantically aware 3DGS pipelines, paving the way for more powerful and generalizable scene understanding in real-world environments.

\begin{acks}
The authors would like to acknowledge the financial support from the National Research Council (NRC) Canada under the grant agreement DHGA AI4L129-2 (CDB \#6835) and from Cognia AI Inc. and Mitacs Inc. through the Accelerate Entrepreneur program.
\end{acks}

\bibliographystyle{ACM-Reference-Format}
\bibliography{references}
\appendix
\end{document}